\documentclass[prx, notitlepage, twocolumn, nofootinbib, superscriptaddress]{revtex4-2}
\usepackage{eurosym}
\usepackage[colorlinks=true,citecolor=blue,linkcolor=blue]{hyperref}
\usepackage{graphicx}
\usepackage{color}
\usepackage{dcolumn}
\usepackage{bm}
\usepackage{mathrsfs}
\usepackage{amsmath}
\usepackage{amssymb}
\usepackage{amsthm}
\usepackage{amsfonts}
\usepackage{enumerate}
\usepackage{latexsym}
\usepackage{hyperref}
\usepackage{centernot}
\usepackage{multirow}

\begin{document}

\title{Revealing Spin and Spatial Symmetry Decoupling: New Insights into
Magnetic Systems with Dzyaloshinskii-Moriya Interaction}
\author{Yuxuan Mu}
\affiliation{National Laboratory of Solid State Microstructures and School of Physics,
Nanjing University, Nanjing 210093, China}
\affiliation{Collaborative Innovation Center of Advanced Microstructures, Nanjing
University, Nanjing 210093, China}
\affiliation{International Quantum Academy, Shenzhen 518048, China}
\author{Di Wang}
\thanks{Corresponding author: diwang0214@nju.edu.cn}
\affiliation{National Laboratory of Solid State Microstructures and School of Physics,
Nanjing University, Nanjing 210093, China}
\affiliation{Collaborative Innovation Center of Advanced Microstructures, Nanjing
University, Nanjing 210093, China}
\author{Xiangang Wan}
\thanks{Corresponding author: xgwan@nju.edu.cn}
\affiliation{National Laboratory of Solid State Microstructures and School of Physics,
Nanjing University, Nanjing 210093, China}
\affiliation{Collaborative Innovation Center of Advanced Microstructures, Nanjing
University, Nanjing 210093, China}
\affiliation{Hefei National Laboratory, Hefei 230088, China}
\affiliation{Jiangsu Physical Science Research Center, Nanjing University, Nanjing
210093, China}

\begin{abstract}
It is widely accepted that spin-orbit coupling (SOC) generally locks spin
and spatial degrees of freedom, as a result, the spin, despite being an
axial vector, is fixed and cannot rotate independently, and the magnetic
system should be described by magnetic space groups (MSGs). While as a new
type of group, spin space groups (SSGs) have been introduced to
approximately describe the symmetry of magnetic systems with negligible SOC,
and received significant attention recently. In this work, we prove that in
two cases of coplanar spin configurations, there are spin-only operations
that strictly hold even with considerable Dzyaloshinskii-Moriya interaction
(DMI), and the symmetry of their spin models could be described by the
spin-coplanar SSG. In addition, we also find that for spin-collinear cases,
regardless the strength of DMI, the magnon systems within the framework of
linear spin wave theory (LSWT) also preserve the decoupled spin and spatial
rotations, but the symmetry does not belong to the conventional definitions
of collinear spin groups. We discuss the potential realization of
these novel symmetries in rod, layer, and
three-dimensional (3D) space groups. 
Our work extends the applicability of SSGs to magnetic materials with heavy elements, 
and reveals that the coexistence of DMI and SSG symmetries provides new opportunity for exploring novel magnon transport phenomena, and 
potential material realization had also been discussed. 
\end{abstract}

\date{\today }
\maketitle

Symmetry, as formulated by group theory, is fundamental to the laws of
nature. In condensed-matter physics, symmetry has long been central, as it
dictates the physical properties, phase transitions, and behavior of various
materials\cite{dresselhaus2007group}. It governs the conservation laws \cite%
{PhysRev.124.287}, determines selection rules \cite{10.1063/1.444442}, and
plays a particularly crucial role in fields of Landau theory of phase
transition \cite{landau2013course}, topological phases \cite%
{RevModPhys.88.035005,PhysRevLett.106.106802,zhang2019catalogue,vergniory2019complete,tang2019comprehensive}
and quasiparticle excitations \cite%
{bradlyn2016beyond,YU2022375,PhysRevB.104.085137}. The symmetries of
three-dimensional (3D) solids are comprehensively described by
crystallographic group theory, with spatial operations, including rotations,
reflections, translations, and their combinations, forming the 32 point
groups (PGs) and 230 space groups (SGs) for nonmagnetic systems \cite%
{bradley2009mathematical}. When spin-orbit coupling (SOC) effect is
considered, the double SGs are applicable. On the other hand, for magnetic
materials, the SOC generally locks spin and spatial degrees of freedom,
preventing spin from rotating independently, and the symmetry of magnetic
systems should be described by 122 magnetic point groups (MPGs) and 1651
magnetic space groups (MSGs) \cite{bradley2009mathematical}.

In 1960s, it is found that the Heisenberg model neglecting the
Dzyaloshinskii-Moriya interaction (DMI), which is the leading order of SOC
effect \cite{DZYALOSHINSKY1958241,PhysRev.120.91},\ enjoys higher symmetries
than MSGs, and the concept of spin space groups (SSGs) is introduced \cite%
{brinkman1966theory,10.1063/1.1708514,LITVIN1974538,Litvin:a14103}. It has
attracted significant attention in areas such as Landau theory of phase
transitions \cite{ssg-add1,ssg-add2}, neutron scattering \cite{ssg-add3},
and magnon band topology \cite%
{PhysRevLett.115.147201,PhysRevLett.119.247202,PhysRevB.105.064430,PhysRevLett.130.206702}%
. Very recently, the full classification of collinear SSGs and the
associated magnon band topology for collinear magnets \cite{chen2023spin}
are developed, offering new pathways to exploring exotic phenomena of
magnonic systems.

In SSGs, the spin and spatial operations become partially decoupled, giving
rise to an expanded set of symmetries expressed as \cite%
{LITVIN1974538,Litvin:a14103,PhysRevX.12.021016},

\begin{equation}
G_{SS}=G_{NSS}\times G_{SO}  \label{1}
\end{equation}%
where $G_{SO}$ stands for the spin-only group that only contains spin
operations, $G_{NSS}$ stands for the nontrivial spin space group that
contains no pure spin operations \cite{PhysRevX.12.021016}.

In recent years, SSGs have been extended to electronic systems, and drawn
growing attention due to their relevance in the study of magnetic materials
with negligible SOC \cite%
{PhysRevX.12.021016,yang2024symmetry,PhysRevLett.127.176401,
PhysRevLett.126.127701,PhysRevLett.128.197202,PhysRevX.12.011028,PhysRevB.99.184432,PhysRevB.102.144441,PhysRevMaterials.5.014409,doi:10.1126/sciadv.aaz8809,PhysRevB.102.014422, PhysRevX.12.040501,PhysRevX.12.031042,PhysRevLett.132.056701,PhysRevLett.132.036702,gomonay2024structure,PhysRevLett.133.056701,bose2022tilted,zhu2024observation,krempasky2024altermagnetic,d41586-024-00190-w,PhysRevLett.134.106801,PhysRevLett.134.106802,zhou2025manipulation,PhysRevX.14.031037,PhysRevX.14.031039,PhysRevX.14.031038}%
. In field of magnetic topological electronic states, the classifications
and the irreducible corepresentations of spin groups for coplanar and
collinear magnetic structures are derived \cite{PhysRevX.12.021016}, and the
Z$_{2}$ topological phases \cite{PhysRevX.12.021016} and exotic nodal-line
or -point semimetals \cite{yang2024symmetry,PhysRevLett.127.176401}
protected by SSG\ symmetry are predicted. Meanwhile, SSGs are also crucial
for understanding the unconventional spin-momentum locking without SOC\cite%
{PhysRevLett.126.127701,PhysRevLett.128.197202,PhysRevX.12.011028,PhysRevB.99.184432,PhysRevB.102.144441,PhysRevMaterials.5.014409,doi:10.1126/sciadv.aaz8809,PhysRevB.102.014422}%
, which was later recognized as a characteristic feature of altermagnetism%
\cite%
{PhysRevX.12.040501,PhysRevX.12.031042,PhysRevLett.132.056701,PhysRevLett.132.036702,gomonay2024structure,PhysRevLett.133.056701,bose2022tilted,zhu2024observation,krempasky2024altermagnetic,d41586-024-00190-w,PhysRevLett.134.106801,PhysRevLett.134.106802,zhou2025manipulation}%
. Very recently, an exhaustive analysis of SSGs symmetries provides a full
classification and investigates the enriched features of electronic bands
within SSGs framework \cite%
{PhysRevX.14.031037,PhysRevX.14.031039,PhysRevX.14.031038}, which paves the
way for deeper understanding and further exploration of emergent phenomena
in magnetic materials.

The symmetry of SSGs is generally an approximation, holding strictly only
when SOC is absent. However, as a relativistic effect, SOC always exists and
plays a crucial role in various fields of condensed matter physics, such as
permanent magnets \cite{campbell1996permanent,skomski2019permanent},
spintronics \cite{SOC-1}, spin-orbitronics \cite{SOC-2}, and topological
physics \cite{SOC-3,RevModPhys.83.1057}. The DMI, which has been widely
regarded as the \textquotedblleft SOC" in spin models and magnon systems
\cite%
{brinkman1966theory,science.1188260,PhysRevLett.117.217202,PhysRevLett.117.217203,magnon-review-1,031620-104715}%
, is also always present even in the centrosymmetric compound, since not all
bond centers are inversion centers. In topological magnons, DMI generates
the Berry curvature and plays a central role in the study of magnon band
topology \cite%
{PhysRevB.87.144101,PhysRevLett.117.157204,PhysRevB.95.224403,PhysRevLett.117.227201,PhysRevApplied.9.024029,PhysRevB.95.224403,PhysRevB.95.014422}%
. Meanwhile, in the field of magnon transport, DMI is essential in many
phenomena such as the magnon Hall effect\cite%
{science.1188260,PhysRevLett.114.047201,PhysRevLett.115.106603,PhysRevB.84.184406,doki2018spin,PhysRevB.91.125413,PhysRevB.99.014427,PhysRevB.98.094419}
and the magnon spin Nernst effect\cite%
{PhysRevLett.117.217203,PhysRevLett.117.217202,PhysRevB.93.161106,PhysRevResearch.2.013079,PhysRevB.97.174407,chen2021observation}%
, where DMI takes on a role analogous to that of SOC in the spin Hall effect.

The DMI and SSG symmetries have both played significant roles in the
exploration of exotic phenomena in magnetic materials. However, as mentioned
above, they are inherently in conflict. Could there be scenarios where SSG
symmetries remain strictly applicable even when huge DMI
is taken into account? The coexistence of DMI and SSG symmetries in one
magnetic system will undoubtedly provide new avenues for study novel
physical phenomena. In this work, we prove that the inclusion of DMI does
not necessarily lead to a complete locking of spin and spatial rotations,
and identify a novel symmetry to describe the spin model and magnon system
with even considerable DMIs.

\textbf{Group Theory Analysis}

First we consider the coplanar spin configurations with the spin moments
lying in the $xy$ plane. The symmetry of spin-coplanar SSGs could be written
as Eq. (\ref{1}), where the spin-only group for spin-coplanar SSGs has the
expresssion as \cite{LITVIN1974538,Litvin:a14103,PhysRevX.12.021016},
\begin{equation}
G_{SO}=\{E,TU_{z}(\pi )\}  \label{7}
\end{equation}%
where $T$ and $U_{z}(\pi )$ represent time-reversal symmetry and the
two-fold spin rotation operation along the z-axis, respectively.

In the spin Hamiltonian, the antisymmetric DMI is expressed as $%
H_{DMI}=\sum_{i,j}\bm{D_{ij}}\cdot (\bm{S_i}\times \bm{S_j})$ \cite%
{DZYALOSHINSKY1958241,PhysRev.120.91}, where $\bm{S_i}$ and $\bm{S_j}$ are the spin operators at the lattice sites $i$ and $j$, while $\bm{D_{ij}}$ is a vector arising from SOC and satisfies the antisymmetry condition $\bm{D_{ij}}=-\bm{D_{ji}}$. 
The spin Hamiltonian may also contain other spin interactions like anisotropic exchange or dipole-dipole couplings, 
but these are typically orders of magnitude weaker than the DM interaction \cite{DZYALOSHINSKY1958241,PhysRev.120.91,PhysRevB.108.085140} and thus are not considered in this work. 
In general, the inclusion of DMI
locks spin and spatial rotations, and reduces the Hamiltonian symmetry to
MSGs. However, we find that under two following cases of magnetic
configurations, the spin and spatial rotations are still partially decoupled:

(i) The first case occurs in 2D systems, when the magnetic atoms lie within
the $xy$ plane, which possesses horizontal mirror symmetry $\sigma _{h}$. In
this case, only the component of DMI perpendicular to the mirror plane, i.e.
$D^{z}$, is left \cite{PhysRev.120.91}, which makes that the spin
Hamiltonian retains the symmetry $\{TU_{z}(\pi )||E\}$. Therefore, the
spin-only group $G_{SO}=\{E,TU_{z}(\pi )\}$ in spin-coplanar SSG is strictly
preserved. Below we provide a criterion to determine whether the operations
in $G_{NSS}$ are preserved:

An operation $R\in G_{NSS}$ in 2D systems could be written as $R=\{A||B\}$, 
where $A$ and $B$ represents the spin and spatial operations, respectively. 
The generators\ for the spin operation $A$\ include $U_z(\frac{2\pi}{n})$ and
 $U_x(\pi)$,where $n=1,2,3,4,6$\cite{PhysRevX.12.021016}, while the generators
  for the spatial operation $B$ include $C_{nz}$, $C_{2x}$, $S_{4}$, $P$, 
  $\sigma _{h}$ and $\sigma _{v}$\cite{litvin2012character}, where $S_{4}$ presents
   the four-fold rotation inversion operation and $P$ presents inversion operation, 
   while $\sigma _{h}$\ and $\sigma _{v}$ correspond to mirror operactions
along the horizontal and vertical planes, respectively. 

The DMI with only $D^{z}$ can be expressed as $H_{DMI}=\sum_{i,j}D_{ij}^{z}(S_{i}^{x}S_{j}^{y}-S_{i}^{y}S_{j}^{x})$. 
By applying the generators to the DMI term, the spatial operations $B$ act on the site $i$ and $j$, 
while the spin operations $A$ act on the directions of spin operations $S^x$ and $S^y$, respectively. 
The spin Hamiltonian is constrained by the space group symmetry and should remain invariant under the space group operation \cite{PhysRev.120.91}.
Note that in the spin group framework, this symmetry operation of space group is expressed as $R_{SG}=\{B'||B\}$, 
where $B'$ denotes the proper rotation part of the spatial operation $B$. 
Therefore, whether a coplanar SSG $R=\{A||B\}$ remains valid depends on whether the two spin-space rotations $A$ and $B'$ alter the sign 
of $S_{i}^{x}S_{j}^{y}-S_{i}^{y}S_{j}^{x}$ in a consistent manner. 
Considering the influence of spin part on the DMI term, we categorize these generators into two types: $S_{i}^{x}S_{j}^{y}-S_{i}^{y}S_{j}^{x}$
changes sign under $U_x(\pi)$ (the corresponding spatial operations $B$ being $C_{2x}$ and $\sigma _{v}$), 
while $U_z(\frac{2\pi}{n})$ leaves $S_{i}^{x}S_{j}^{y}-S_{i}^{y}S_{j}^{x}$ invariant 
(the corresponding spatial operactions $B$ are $C_{nz}$, $S_{4}$, $P$ and $\sigma _{h}$).
For a given operation $R=\{A||B\}$, 
when the operations $A$ and $B$ belong to the same type described above,
the operation $R$ is preserved even in the presence of DMI. Otherwise, the
operation $R$ would be broken.

(ii) The second case arises in 1D systems, when the magnetic atoms are
aligned along $z$-axis chain, and the chain has double rotational symmetry $%
C_{2z}$, which also constrain the direction of the DMI to $z$ direction\cite%
{PhysRev.120.91}. Therefore, the DMI with only $z$-components $D_z$ ensures that the SSG symmetry  $\{TU_{z}(\pi )||E\}$ is preserved. Analogous to case (i), the symmetry of spin Hamiltonian maintain the form of spin-coplanar SSG with the spin-only group $G_{SO}=\{E,TU_{z}(\pi )\}$. For any given SSG operation $R=\{A||B\}$, one can also determine whether it remains valid using the criterion discussed in case (i).

\bigskip

Furthermore, for collinear spin configuration within the framework of the
linear spin wave theory (LSWT), we also find interesting symmetries here.
The spin-only group $G_{SO}$ for the collinear spin configurations could be
written as \cite{LITVIN1974538,Litvin:a14103,PhysRevX.12.021016},
\begin{equation}
G_{SO}=\{E,TU_{n}(\pi )\}\ltimes U_{z}(\phi )  \label{9}
\end{equation}%
where the direction of spin moments is defined along the $z$-direction, and
$U_{z}(\phi ),$ $\phi \in \lbrack 0,2\pi )$ means full spin rotation
along $z$-axis, $\{E,TU_{n}(\pi )\}$ contains the identity operation as well
as the combined operation of time reversal $T$ and two-fold spin rotation
about $n$-axis perpendicular to the spin $z$-axis. Within the framework of
LSWT, the Holstein-Primakoff (HP) transformation results in that only $%
D_{ij}^{z}$ term contains quadratic contribution of magnons. Through
symmetry analysis, we demonstrate that the $D_{ij}^{z}$ term preserves $%
\{U_{z}(\phi )||E\}$ symmetry but breaking the operation $\{TU_{n}(\pi
)||E\} $. Therefore, the magnons in spin-collinear systems has the novel\
symmetry expressed as $G_{SS}=G_{NSS}\times G_{SO}^{\prime }$, where the
spin-only group could be written as
\begin{equation}
G_{SO}^{\prime }=U_{z}(\phi )
\end{equation}%
while the nontrivial spin space group $G_{NSS}$ is a subgroup of the one for
original SSG without DMI. It should be noted that, the above expression of
the spin group $G_{SS}$ has spin-only symmetry beyond MSG, but does not
belong to the conventional definitions of collinear or coplanar spin groups,
whose spin-only groups are given by Eq. (\ref{7}) and (\ref{9}) respectively.

\bigskip

The coexistence of DMI interactions with SSG symmetries directly gives rise to emergent physical phenomena. 
In the following we present a concrete application in magnon transport.
In magnon spintronics, the ability to generate a pure
transverse spin current without a thermal current is of great interest \cite%
{PhysRevLett.117.217202,PhysRevLett.117.217203}.  
Generally, the DM interactions play an important role in the magnon spin Hall effect \cite{science.1188260,PhysRevLett.117.217202,PhysRevLett.117.217203,PhysRevLett.114.047201,PhysRevLett.115.106603,PhysRevB.84.184406,doki2018spin}, 
thus researchers typically employ MSG symmetries to analyze/discuss the system with magnon spin Hall but without magnon thermal Hall.
For example, in Refs \cite{PhysRevLett.117.217202,PhysRevLett.117.217203}, the suppression of the magnon thermal Hall effect 
is achieved in collinear antiferromagnets via $PT$ symmetry, where $P$ and $T$ represent the inversion and time-reversal symmetry. 
The using of MSG significantly narrows the range of materials capable of generating pure magnon spin currents. 

In contrast, in the two spin-coplanar cases we proposed, the SSG symmetry $\{TU_{z}(\pi )||E\}$ could ensure that the magnon Berry curvature 
is an odd function with respect to the quasimomentum $\bm{k}$, $\Omega _{n}(\bm{k})=-\Omega _{n}(-\bm{k})$, 
thereby completely suppressing the magnon thermal Hall response. 
Notably, the DMI of $D^{z}$ in these cases induces 
the magnon spin conductivity $\sigma^{z}_{\alpha \beta}$,  
where the in-plane indices $\alpha$ and $\beta$ denote the directions of the magnon spin current and the temperature gradient, respectively. 
Thus, the pure magnon spin current driven by a temperature gradient can be achieved in systems with the coexistence of DMI interactions and SSG symmetries. 
This is different from previous theoretical frameworks \cite{PhysRevLett.117.217202,PhysRevLett.117.217203}.
In the following, we would also present the potential material realization.

\begin{table}[tbp]
\caption{We list all 17 layer groups with mirror symmetry $\protect\sigma %
_{h}$, if the magnetic atoms of 2D materials occupy the given Wyckoff
positions (i.e. $z$=0)\ and their spins lie within the xy plane, the spin
Hamiltonian including DMI maintain the symmetry of spin-coplanar SSG,
written as Eq. (\protect\ref{1}) and Eq. (\protect\ref{7}).}
\label{table2}\centering
\begin{ruledtabular}
        \begin{tabular}{lll}
            \textbf{Layer Group} & \textbf{Wyckoff Positions} \\ \hline
            $p11m$              & 1a                         \\
            $p112/m$            & 1a,1b,1c,1d,2i             \\
            $p m 2 m$              & 1a,1b,2e                  \\
            $p b 2_1  m$          & 2a                         \\
            $c m 2 m$              & 2a,4c                      \\
            $p m m m$              & 1a,1b,1c,1d,2e,2f,2g,2h,4q \\
            $p m a m$              & 2a,2b,2c,4f                \\
            $p b a m$              & 2a,2b,4e                  \\
            $c m m m$              & 2a,2b,4c,4d,4e,8k          \\
            $p 4/m$                & 1a,1b,2c,4g                \\
            $p 4/m m m$            & 1a,1b,2c,4g,4h,4i,8j       \\
            $p 4/m b m$            & 2a,2b,4e,8f                \\
            $p-6$                  & 1a,1b,1c,3g                \\
            $p 6/m$                & 1a,2b,3d,6g                \\
            $p-6 m 2$              & 1a,1b,1c,3g,6h             \\
            $p-6 2 m$              & 1a,2b,3d,6g                \\
            $p 6/m m m$            & 1a,2b,3d,6g,6h,12k
            \end{tabular}
    \end{ruledtabular}
\end{table}

\begin{table}[tbp]
\caption{The selected 2D candidate materials from C2DB \protect\cite%
{Haastrup_2018,Gjerding_2021}, with magnetic atoms occupying given Wyckoff
positions in TABLE \protect\ref{table2} and exhibiting in-plane spin
configurations.}
\label{table4}\centering
\begin{ruledtabular}
    \begin{tabular}{llllll}
    \textbf{Layer Group}  & \textbf{Materials} \\ \hline
    $p m 2 m$   &   $\mathrm{NiCl_2N_2H_4C_6}$,$\mathrm{MoV_3Se_8}$,               \\
                &   $\mathrm{WV_3Se_8}$,$\mathrm{MoV_3Te_8}$,                       \\
                &   $\mathrm{WV_3Te_8}$,$\mathrm{OVBr_2}$,              \\
                &   $\mathrm{OVCl_2}$,$\mathrm{OVF_2}$,                \\
                &   $\mathrm{OVI_2}$,$\mathrm{Mo_2V_2Se_8}$,              \\
                &   $\mathrm{Mo_2V_2Te_8}$,$\mathrm{V_2W_2Te_8}$                \\
    $c m 2 m $  &   $\mathrm{FeOBr_2}$                               \\
    $p m m m$   &   $\mathrm{CrOF_2}$                       \\
    $c m m m $  &   $\mathrm{OsVO_2Br_4}$        \\
    $p 4/m m m$ &   $\mathrm{Co_2S_2}$,$\mathrm{Ni_2S_2}$             \\
    $p-6 m 2 $  &   $\mathrm{VS_2}$,$\mathrm{VSe_2}$,$\mathrm{VTe_2}$,                    \\
                &   $\mathrm{YBr_2}$,$\mathrm{YI_2}$,$\mathrm{FeBr_2}$,         \\
                &   $\mathrm{FeI_2}$,$\mathrm{MnTe_2}$,$\mathrm{NbTe_2}$,                   \\
                &   $\mathrm{ScBr_2}$,$\mathrm{ScCl_2}$,$\mathrm{ScI_2}$,                     \\
                &   $\mathrm{ScSe_2}$,$\mathrm{CrAs_2}$,$\mathrm{CrBi_2}$             \\
    $p-6 2 m $  &   $\mathrm{Ti_2I_6}$
    \end{tabular}
    \end{ruledtabular}
\end{table}

\textbf{Candidate Materials}

Based on the above symmtry analysis, we will extract the criteria to
identify the magnetic materials whether the spin and spatial operations are
still partially decoupled even with large DMI. The 2D systems are described
by wallpaper groups or layer groups \cite{hahn1983international}. In the
case of wallpaper groups, only the $xy$-direction is defined, thus the $xy$%
-plane is inherently mirror symmetric. For all the 17 wallpaper groups, when
the spins lie in the $xy$-plane, the spin-only group $G_{SO}$ in
spin-coplanar SSG is preserved.

\begin{table*}[tbp]
    \caption{ We provide the symmetry criteria applicable to 3D magnetic
    systems by mapping 2D layer groups to 3D space groups as discussed in the
    main text. In these space groups, when magnetic atoms occupy given Wyckoff
    positions with coplanar magnetic moments, and the interlayer distance is
    relatively large, the huge DMI does not degrade the spin Hamiltonian
    symmetry to MSGs, but preserve the expression of SSG symmetry with spin-only
    group as Eq. (\protect\ref{7}).}
    \label{table6}\centering
    \begin{ruledtabular}
            \begin{tabular}{llllll}
                 \textbf{Layer Group}   & \textbf{Space Group}   & \textbf{Wyckoff Positions(Space Group)} \\ \hline
                $p 1 1 m $              & $Pm$     & 1a,1b         \\
                $p 1 1 2/m$             & $P2/m$   & 1a-1h,2m,2n        \\
                $p m 2 m$               & $Pmm2$     & 1a-1d,2e-2f                  \\
                $p b 2_1  m$            & $Pmc2_1$     & 2a,2b                      \\
                $c m 2 m$               & $Amm2$     & 2a,2b,4d,4e                     \\
                $p m m m$               & $Pmmm$     & 1a-1h,2i-2p,4y,4z           \\
                $p m a m $              & $Pmma$     & 2a-2f,4i,4j              \\
                $p b a m$               & $Pbam$         & 2a-2d,4g,4h                \\
                $c m m m $              & $Cmmm$      & 2a-2d,4e-4j,8p,8q                \\
                $p 4/m$                  & $P4/m$      & 1a-1d,2e,2f,4j,4k          \\
                $p 4/m m m$             & $P4/mmm$      & 1a-1d,2e,2f,4j-4o,8p,8q    \\
                $p 4/m b m$              & $P4/mbm$    & 2a-2d,4g,4h,8i,8j    \\
                $p-6 $                   & $P-6$    & 1a-1f,3j,3k     \\
                $p 6/m$                  & $P6/m$ & 1a,1b,2c,2d,3f,3g,6j,6k     \\
                $p-6 m 2$                & $P-6m2$          & 1a-1f,3j,3k,6l,6m      \\
                $p-6 2 m$                & $P-62m$      & 1a,1b,2c,2d,3f,3g,6j,6k    \\
                $p 6/m m m$             & $P6/mmm$        & 1a,1b,2c,2d,3f,3g,6j-6m,12p,12q
            \end{tabular}
            \end{ruledtabular}
    \end{table*}

    \begin{table*}[tbp]
    \caption{We list the 1D rod groups with twofold rotational symmetry $C_{2z}$
    along the $z$-axis. With spins lying in the $xy$ plane, the magnetic
    materials that meet the symmetry criteria outlined in the table will exhibit
    additional SSG symmetry.}
    \label{table3}\centering
    \begin{ruledtabular}
            \begin{tabular}{llllll}
                \textbf{Rod Group}   & \textbf{Wyckoff Positions}  & \textbf{Rod Group}   & \textbf{Wyckoff Positions} \\ \hline
                $p 1 1 2$     & 1a                  & $p-4 2 m$      & 1a,1b,2c          \\
                $p 1 1 2/m$   & 1a,1b,2c           & $p-4 2 c$      & 2a,2b,4e          \\
                $p 2 2 2$     & 1a,1b, 2g          & $p 4/m m m$    & 1a,1b,2c         \\
                $p m m 2$     & 1a                  & $p 4/m c c$    & 2a,2b,4c          \\
                $p c c 2$     & 2a                  & $p 4_2/m m c$ & 2a,2b,4c          \\
                $p m m m$     & 1a,1b,2g          & $p 6$          & 1a                \\
                $p c c m$     & 2a,2b,4e          & $p 6_2$       & 3a                \\
                $p 4$         & 1a                 & $p6_4$        & 3a                \\
                $p 4_2$      & 2a                  & $p 6/m$        & 1a,1b,2c          \\
                $p-4$         & 1a,1b,2c            & $p 6 2 2$      & 1a,1b,2c          \\
                $p 4/m$       & 1a,1b,2c            & $p 6_2  2 2 $ & 3a,3b,6c          \\
                $p 4_2/m$    & 2a,2b,4c            & $p 6_4  2 2$  & 3a,3b,6c          \\
                $p 4 2 2$     & 1a,1b,2c            & $p 6 m m$      & 1a                \\
                $p 4_2 2 2$       & 2a,2b,4c            & $p 6 c c$      & 2a                \\
                $p 4 m m$     & 1a                  & $p 6/m m m$    & 1a,1b,2c          \\
                $p 4_2 c m$     & 2a                  & $p 6/m c c$    & 2a,2b,4c          \\
                $p 4 c c $    & 2a                 &              &
            \end{tabular}
            \end{ruledtabular}
    \end{table*}

On the other hand, since most 2D materials typically consist of multiple
atomic layers or have finite extension in the $z$-direction, the layer group
usually provides a more appropriate description \cite{hahn1983international}%
. There are 17 layer groups having the mirror symmetry $\sigma
_{h} $, which are summarized in TABLE \ref{table2}. If a magnetic material
belongs to the layer groups in TABLE \ref{table2}, and its magnetic atoms
are also located at the given Wyckoff positions in TABLE \ref{table2}, i.e. $%
z=0$, while their spins lie in the $xy$-plane, the presence of DMI does not
degrade the spin Hamiltonian symmetry to MSGs, but instead preserves the
spin-coplanar SSGs symmetries. We predict that the coexistence of
DMI and SSG symmetries exists in 17 out of 80 layer groups ($21\%$), 
demonstrating that this phenomenon is not limited to isolated cases but rather occurs widely across diverse magnetic systems.

For any 2D magnetic material,
one can easily identify their symmetries using our TABLE \ref{table2}.
To illustrate the result, we apply our theory to a popular database Computational 2D Materials Database (C2DB)\cite{Haastrup_2018,Gjerding_2021}
and identify 33 materials as summarized in TABLE \ref{table4}.
The spin Hamiltonian for these materials would maintain the SSG symmetry even with large DMI. 
As an illustrative example, we consider the layered material $\mathrm{VSe_2}$ in TABLE \ref{table4}, whose layer group is $p-6m2$, with magnetic V atoms residing on the mirror plane. 
It was suggested that $\mathrm{VSe_2}$ adopts an in-plane collinear ferromagnetic ground state \cite{fuh2016newtype}. 
Due to the strong SOC in selenium \cite{zhang2009topological}, 
this material would exhibit significant DM interactions. 
Conventionally, such materials should be described by MSG theory, 
thus naively one will expect both magnon thermal Hall and the magnon spin conductivity. 
However, our symmetry analysis above reveals that the DMI and the SSG symmetry $\{TU_{z}(\pi )||E\}$ 
in this compound induces magnon spin Hall and forbids the magnon thermal Hall, respectively. 
Thus, the coexistence of DMI and SSG symmetry results in the pure magnon spin current ($\sigma^{z}_{\alpha \beta}$) in this material.

In addition to the classifications in 2D groups in TABLE \ref{table2}, we
can also expand it to 3D systems through the corresponding mappings from
layer groups to 3D space groups \cite{litvin2012character}. Each layer group
is isomorphic to the factor group $G/T(1)$, where $G$ represent a 3D space
group, and $T(1)$ is a 1D translation subgroup of $G$. It allows each layer
group to correspond to a specific 3D space group, which represents the
structure obtained when the layer group is extended along the $z$ direction.
The corresponding space groups and required Wyckoff positions of the
magnetic atoms for 3D materials are summarized in TABLE \ref{table6}. For
3D materials with sufficiently large spacing between their mirror
planes, where interlayer anisotropic interactions can be neglected as an additional approximation, 
one can also determine whether they possess decoupled spin and spatial
symmetry using TABLE \ref{table6}.

For 1D materials, their symmetry is typically described using rod groups,
which possess translational symmetry along a single axis (marked as $z$%
-direction) with 3D space\cite{hahn1983international}. Based on the symmetry
analysis above, we provide the TABLE \ref{table3}, when magnetic materials
belongs to the given rod groups and the magnetic atoms occupy the required
Wyckoff positions, while their magnetic moments lie in the $xy$ plane,\
these materials could preserve spin-coplanar SSG symmetries described by Eq.
(\ref{1}) and Eq. (\ref{7}). The TABLE \ref{table3} can also be expanded to
identify the quasi-1D magnetic candidate materials. Since many magnetic
materials exhibit quasi-1D characteristics, such as Sr$_{3}$MIrO$_{6}$(M =
Li, Na, Zn, Cd) \cite{segal1996structural}, thereby the exploration of SSG
symmetry in these materials is worthy of further investigation.

\textbf{Acknowledgements}

This work was financially supported by the National Natural Science
Foundation of China (Grants No. 12188101, No. 12334007, and No. 12474233),
the National Key R\&D Program of China (Grant No. 2022YFA1403601),
Innovation Program for Quantum Science and Technology (Grant
No.2021ZD0301902, and No. 2024ZD0300101), Natural Science Foundation of
Jiangsu Province (No. BK20233001, BK20243011).

\clearpage
\bibliographystyle{aps}
\bibliography{2024}

\end{document}